\documentclass[twocolumn]{aastex7}

\usepackage{amsmath,amssymb}

\newcommand{\iras}{IRAS 15398$-$3359}
\newcommand{\hhco}{H$_{2}$CO $(3_{0,3}-2_{0,2})$}

\newcommand{\kms}{km s$^{-1}$}

\newcommand{\twelveco}{{}$^{12}$CO $(J=2-1)$}
\newcommand{\cs}{CS~$(J=5-4)$}

\received{March 31, 2025}
\revised{May 8, 2025}
\accepted{May 12, 2025}

\submitjournal{The Astrophysical Journal}

\begin{document}

\title{Origin of the Shell Structure in the Primary Outflow from IRAS 15398-3359}

\author[0000-0002-7538-581X]{Tomoyuki Hanawa}
\affiliation{Center for Frontier Science, Chiba University,
1-33 Yayoi-cho, Inage-ku, Chiba 263-8522, Japan}
\email[show]{hanawa@faculty.chiba-u.jp}

\author[0000-0003-3655-5270]{Yuki Okoda}
\affiliation{NRC Herzberg Astronomy and Astrophysics, 5071 West Saanich Road, Victoria, BC, V9E 2E7, Canada}
\affiliation{Star and Planet Formation Laboratory, RIKEN Cluster for Pioneering Research, Wako-shi, Saitama, 351-0106, Japan}
\email{Yuki.Okoda@nrc-cnrc.gc.ca}

\author[0000-0001-8227-2816]{Yao-Lun Yang}
\affiliation{Star and Planet Formation Laboratory, RIKEN Cluster for Pioneering Research, Wako-shi, Saitama, 351-0106, Japan}
\email{yaolunyang.astro@gmail.com}

\author[0000-0002-3297-4497]{Nami Sakai}
\affiliation{Star and Planet Formation Laboratory, RIKEN Cluster for Pioneering Research, Wako-shi, Saitama, 351-0106, Japan}
\email{nami.sakai@riken.jp}

\begin{abstract}
IRAS 15398-3359, a Class 0 protostar in Lupus I star forming region, is associated with three 
generations of outflows. The primary outflow, i.e., the most recent one, shows internal structure 
named ``shell structure'' in the near infrared emission map.  The shell structure
is also seen in the emission lines of CO, H$_2$CO, and others species.  We find a similar structure in
an underexpanded jet produced in aerodynamics and other engineering applications. A high pressure gas ejected 
through a nozzle expands to form a supersonic flow. When the pressure of the ejected gas becomes
lower than that of the ambient gas, the jet is compressed to form a shock wave.
The shock heated gas expands again to form substructures along the jet.
We examine the similarity between the primary outflow of IRAS 15398-3359 and industrial 
underexpanded jet and the possibility that the shell structure of the former is due to repeated
expansion and compression in the direction perpendicular to the jet propagation.
\end{abstract}


\section{Introduction}

Jets and outflows are important signatures of protostars. They emerge within 
100~au from a protostar and can reach a few parsec away from the origin.
Recent high resolution imaging observations with ALMA and JWST are valuable 
for studying their launch.  \iras\ is one of the sources that 
both ALMA and JWST have observed to obtain sub-arcsecond images in the radio and infrared wavelengths.

\iras\ is a Class 0 protostar in the Lupus I star forming region located at a distance of 154.9$^{+3.2}_{-3.4}$ pc \citep{galli20} and
is associated with multiple outflows. \cite{tachihara96} discovered a CO outflow with a 4m single dish telescope. 
\cite{oya14} observed the primary outflow extended in the northeast and southwest with ALMA. The northeastern and southwestern outflows are red- and blue-shifted respectively. \cite{2021ApJ...910...11O} discovered a secondary outflow extending to the southeast. The secondary outflow is 8$^{\prime\prime}$ away from the protostar on the plane of the sky, while the primary outflow seems to be connected to the protostar. \cite{sai24} discovered another pair of outflows at a distance of $30 ^{\prime\prime} - 75 ^{\prime\prime}$ in the northern and southern regions.  
This third outflow is also reported by \cite{guzman24}.
The secondary and third outflows indicate that the outflow activity of \iras\ is episodic and the primary outflow is due to the most recent activity \citep{2021ApJ...910...11O,sai24,guzman24}. Since these three generations of outflow extend in different directions, they are likely to be driven by different alignment channels.

The primary outflow has an internal structure. The integrated intensity of \hhco\ emission line shows a node at a distance of $\sim 5^{\prime\prime} $ in the blue-shifted outflow \citep[see Figure 1(a) of][]{2021ApJ...910...11O}. \cite{vazzano21} shows the change in the intensity-weighted line-of-sight velocity in \twelveco\ near this node. The infrared image taken with JWST shows four shells also in the blue-shifted outflow \citep{yang22}.  Figure \ref{MIRI-RGB}, the reproduction of Figure 1 of \cite{okoda25}, shows the infrared emission obtained with the three filters (F560W, F770W, and F1000W) of JWST by a composite image. 
The base of the primary outflow is bright in infrared. Since the infrared emission is rich in the rotational transition of H$_2$ molecules \citep{yang22}, the outflow gas is hot ({$ T \ga 10 ^3~{\rm K}$) at least within the regions of several hundreds au \citep[see][for the temperature distribution]{okoda25}. The thermal energy may accelerate the outflowing gas.  Since the internal structure seen in the infrared emission is correlated with that observed in the radio line emissions, they are likely to have a common origin. 

\begin{figure}
    \centering
    \includegraphics[width=0.5\textwidth]{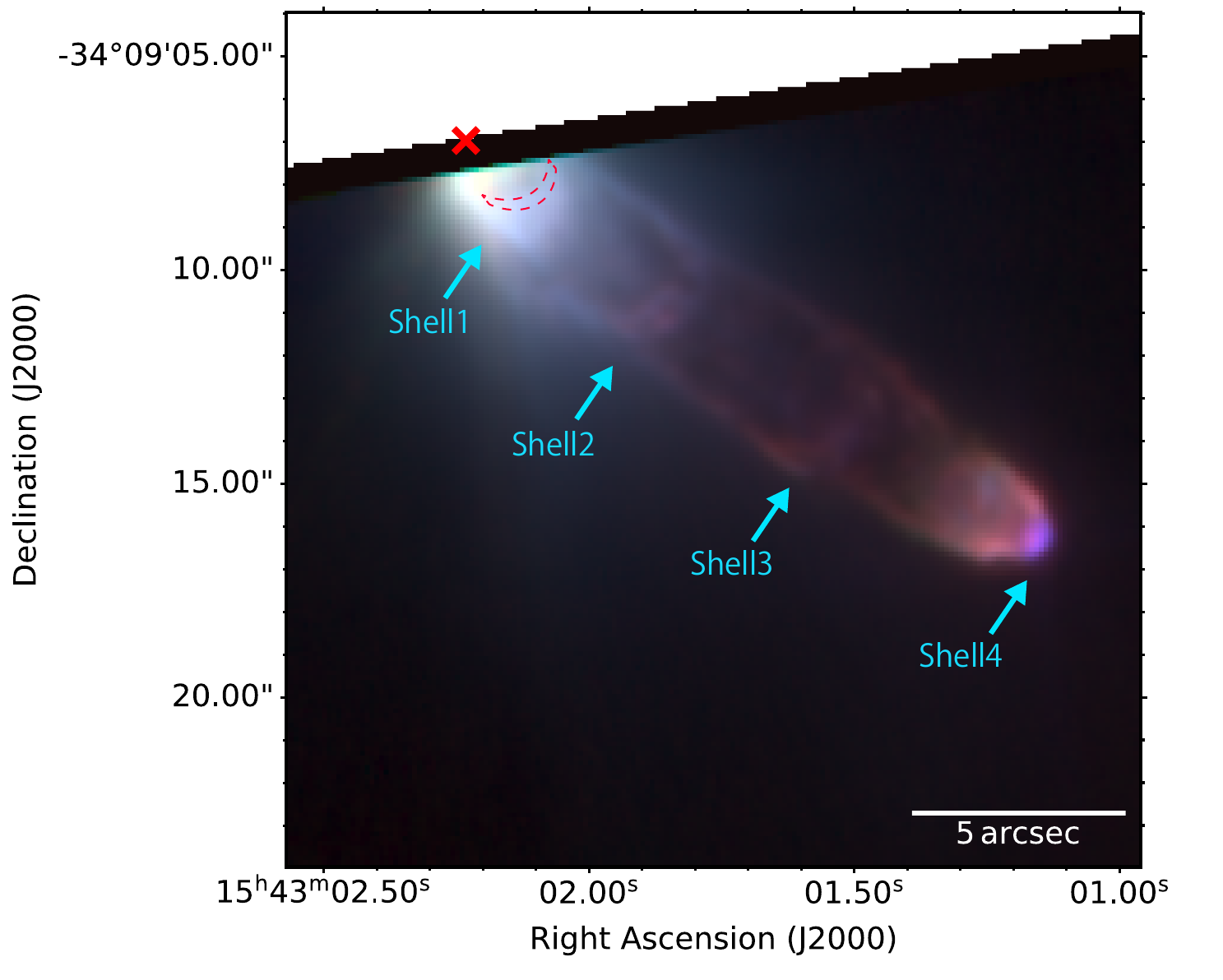}
    \caption{RGB image built from a composite of the three filter (F560W, F770W, and F1000W) MIRI images. The angular separation 5$^{\prime\prime}$ correspond to the deprojected distance of 824 au for the assumed inclination angle of 20$^\circ$ \citep{oya14}. Reproduction of the top panel of Figure 1 of  \cite{okoda25}.\label{MIRI-RGB}}
\end{figure}

A possible origin of the internal structure is time variability of the outflow activity. \cite{vazzano21} considered a scenario that \iras\ has undergone four accretion outbursts which have initiated four times of outflow activities. Another possibility is magnetohydrodynamical instability of the outflow \citep[see, e.g.,][]{shang23}. However, these scenarios are not conclusive. It is worth seeking another possibility as an origin of the internal structure in the primary outflow. 

Jets and outflows have been studied for more than a half century in aeronautical engineering \citep[see, e.g,][for a review]{franquet15}. A high pressure gas emerging from a nozzle evolves into a supersonic flow called ``underexpanded outflow.''  Figure~\ref{fig:barrel-fig} shows the internal structure of the underexpanded outflow for a high jet/ambient over pressure ratio schematically. For simplicity, we omitted expansion fans often included in the similar diagrams. The initially high pressure gas expands preferentially in the direction normal to the exit plane of the nozzle to be collimated.  It is well known that the the underexpanded jet outflow has ``barrel structure'', i.e., semi-periodic change in the outflow width. The barrel closest to the nozzle is bounded by shock waves (see Figure \ref{fig:barrel-fig}).  The thick blue line is called Mach disk where the gas is highly compressed through a shock. The side of the barrel is bounded by oblique shocks.  The jet gas is decelerated once to be hot, dense, and subsonic at the shock. The gas decelerated at the Mach disk expands again to be accelerated and repeats expansion from and convergence to the jet axis. We think that a similar mechanism is responsible for the shell structure of \iras. 

\begin{figure}
\plotone{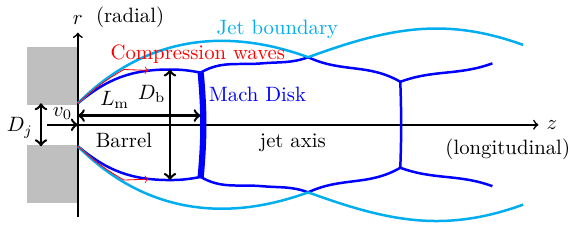}
\caption{A schematic view of an underexpanded outflow from a nozzle.  The blue curves denote the shock waves and highlighted by the thick curve is the Mach disk. The cyan curves denote the jet boundary, i.e., the boundary between the outflowing and ambient gases. The symbols and letters explain the terminology used in the text.  \label{fig:barrel-fig}}
\end{figure}

In this paper we compare \iras\ and engineering underexpanded jet outflow and its substructure to examine the possibility that the former's shell structure and latter's barrel structure are formed by the same physical mechanism. If they have a common origin, we can repurpose the knowledge in industrial science for our understanding of protostellar outflows.

This paper is organized as follows. We analyze the outflow from \iras\ using the archival data of ALMA and JWST in \S 2. We review the underexpanded jet outflow in \S 3 and show a simple hydrodynamic model in \S 4.  Discussions and conclusion are given in \S 5.

\section{Observations}

\subsection{Observation Specifications}

Our analysis is based on the archival data of ALMA and JWST.
The \hhco\ and \cs\ lines were observed in the project 2018.1.01205.L (PI: Satoshi Yamamoto), as a part of ALMA large program FAUST \citep[][\href{http://faust-alma.riken.jp}{http://faust-alma.riken.jp}]{codella21}. The \twelveco\ line was taken from the ALMA archival data \citep[2013.1.00879.S;][]{2013ApJ...772...22Y}. 
The observation parameters are described by \cite{2021ApJ...910...11O}, \cite{2023ApJ...948..127O}, and \cite{2017ApJ...834..178Y}.
The data reduction details of both data are also given by \cite{okoda25}. 

The MIRI images and spectral cubes were taken with the Medium Resolution Spectroscopy (MRS) by the JWST CORINOS program \citep{yang22} (program 2151, PI: Y.-L. Yang).  The imaging data were reduced with JWST calibration pipeline v.1.14.0 \citep{2024zndo..10870758B} using the calibration reference data \texttt{jwst\_1231.pmap}, while the spectroscopic data were reduced with JWST calibration pipeline v.1.12.5 \citep{2023zndo..10022973B} using the calibration reference data \texttt{jwst\_1183.pmap}.  The MIRI images were taken with three filters, F560W, F770W, and F1000W, centered at 5.6, 7.7, and 10.0 \micron, respectively.  The FWHM of the PSF ranges from 0\farcs{207} to 0\farcs{328}. We use the imaging data taken simultaneously with the spectroscopic observation pointing toward a dedicated background position, because the field of view (FoV) of the image serendipitously covers the blue-shifted outflow of \iras.  The MRS data cover the 4.9--28 \micron\ range with four integral field units, which have their FoVs range from 3\farcs{2}$\times$3\farcs{7}\ to 6.6\arcsec$\times$7.7\arcsec.  The spectral resolving power ($\lambda/\Delta\lambda$) decreases from $\sim3700$ to $\sim1300$ with wavelengths.

\subsection{Observation Results}

Figure \ref{MIRI-RGB} shows the southwest part of the primary outflow by RGB image built from a composite of three filter (F560W, F770W, and F1000W) MIRI images \citep{okoda25}. The MIRI image shows four bright shells named shells 1, 2, 3, and 4 in the ascending distance from the protostar.
As mentioned earlier, the MIRI image traces a hot molecular gas since the line emission of H$_2$ molecule is dominant in the filters \citep[see, e.g,][]{yang22}. The shell structure implies the spatial variation in the temperature and density.
The first shell is $d_{\rm s1} = $340 au (2\farcs{06} from the protostar on the sky) while the diameter of the barrel is $ D_{\rm w1} = $ 322~au (2\farcs{08}). The aspect ratio, $ d_{\rm s1} / D_{\rm w1} = 1.05 $, gives a clue to estimate the pressure ratio when an underexpanded outflow model is applied. The second shell is $ d_{\rm s2} = $ 1060~au (6\farcs{42}) away from the protostar and the jet diameter is $ d_{\rm w2} = $ 448~au (2\farcs{87}).  Here,  ${d _{\rm s1}}$ and  ${d _{\rm s2}}$ denote the deprojected distances while $ D_{\rm w1} $ and $ D_{\rm w2} $ denote the projected distance on the sky. The aspect ratios derived from these numbers are also useful for assessing our model.

\begin{figure*}[ht]
    \centering
    \includegraphics[width=\textwidth]{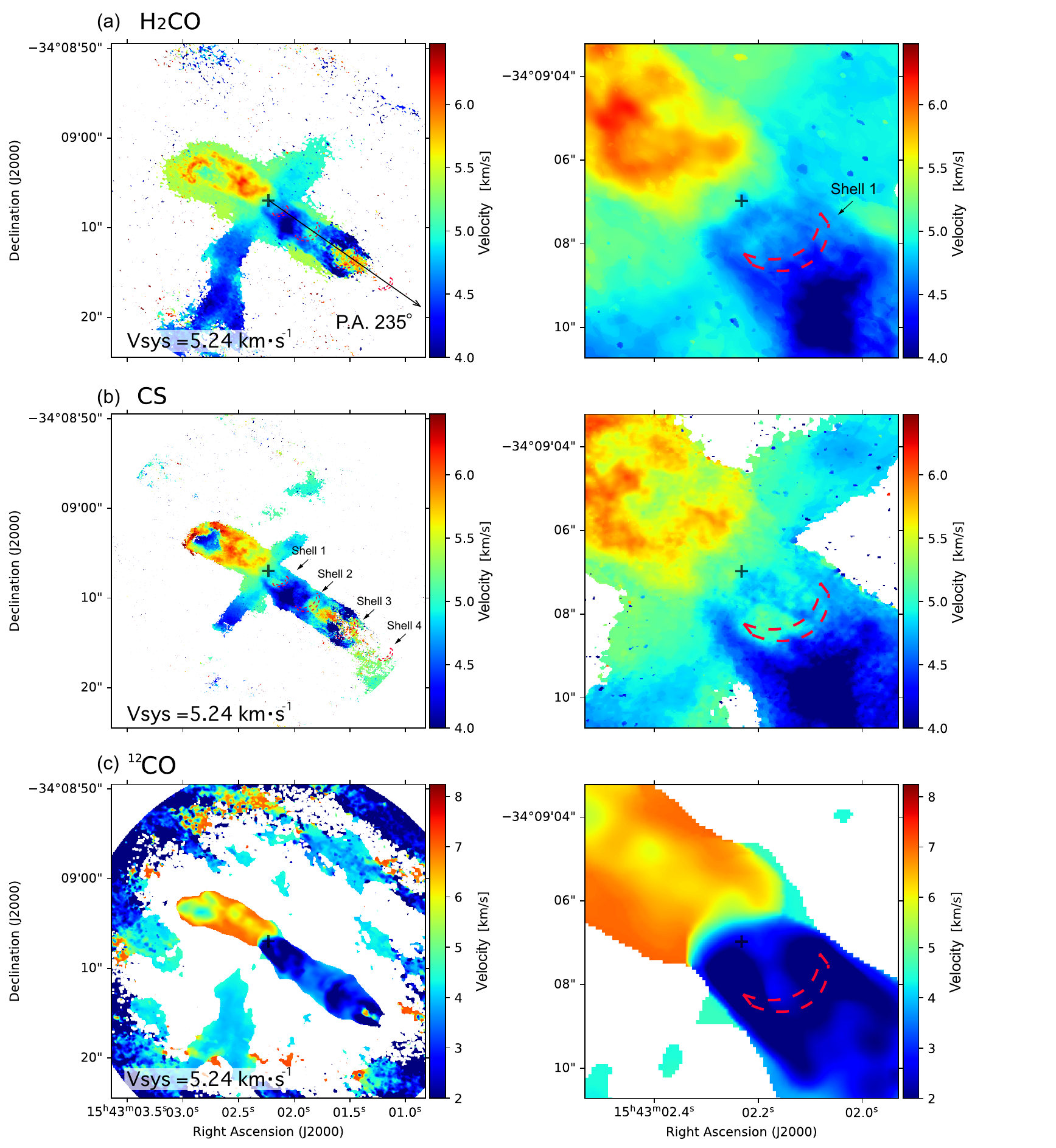}
    \caption{Moment 1 maps of (a) \hhco\, (b) \cs\, and (c) \twelveco. The right panels are the enlargement of the left panels around the protostar ($+$). The jet axis (P.A. 235\degr) and the shell locations are shown in the left panels of (a) and (b), respectively. The red dashed curves in the right panels represent shell 1 shown in Fig.~\ref{MIRI-RGB}. 
    \label{fig:moment1}}
\end{figure*}

\begin{figure}[ht]
    \centering
    \includegraphics[width=0.45\textwidth]{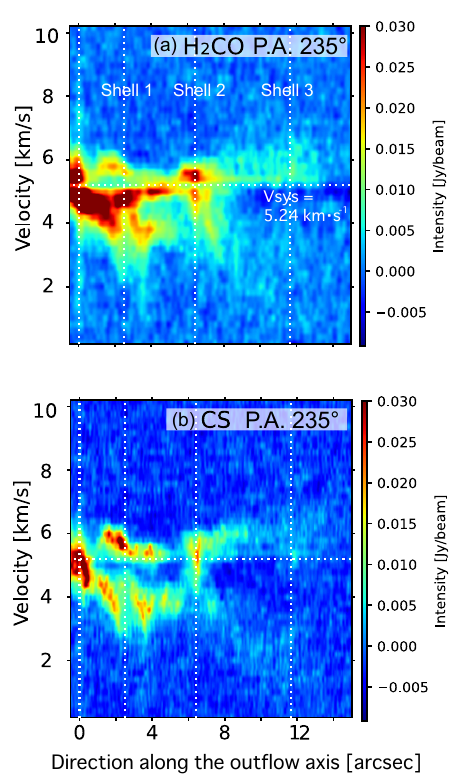}
\caption{PV diagrams along the outflow axis (P.A. 235 \degr). The white dotted lines show the positions of the four shell structures. The systemic velocity of 5.24 \kms\ is shown in the horizontal dotted lines. 
    }
    \label{fig:pv}
\end{figure}

A similar structure is also seen in the molecular emission lines of \hhco, \cs\ \citep{okoda21}, and \twelveco\ \citep{vazzano21}.
Not only the integrated intensity but also the Doppler shift of molecular line emission shows features correlated with the shell structure \citep[see Figure 4 of][]{vazzano21}. Throughout this paper we adopt the systemic velocity of $ v _{\rm sys} = 5.24$~\kms, the central value measured by \cite{2017ApJ...834..178Y} from the ${\rm C}^{18}{\rm O}$ ($J=$2$-$1) emission line.  Figure~\ref{fig:moment1} shows the weighted velocity distribution and position-velocity of (a) \hhco, (b) \cs, and (c) \twelveco\  emission lines.  The left panels show the whole primary outflows while the right panels are enlargement around the protostar.  The position of the protostar is designated by the plus signs on each panel. The locations of shells 1, 2, 3, and 4 are shown in the left panel of Figure~\ref{fig:moment1}b.  Shell 1 is denoted also by the red dashed curves in the right panels. We find a semi-periodic change in the velocity not only in the blue lobe but also in the red lobe.  See also Figure 4 of \cite{thieme23} for a higher angular resolution map of \twelveco\ around the protostar. The weighted velocity of \hhco\ and \cs\ are similar while that of \twelveco\ is slightly different.  We think that the difference is mainly due to the optical depth, i.e., the abundance and excitation condition.  The velocities of \hhco\ and \cs\ show only small deviation from the systemic velocity ($ v _{\rm sys} =$ 5.24 \kms) around the protostar.  The line emission is clearly blue-shifted between shells 1 and 2. This velocity change implies acceleration of the outflow.  The weighted velocity between shells 1 and is low especially in the \cs\ emission.  Note the red-shifted emission in the blue lobe and the blue-shifted one in the red lobe \citep[see also Figure 2 of][]{okoda25}.  These apparently reverse flows appear near the outflow axis.  The slow and reverse flows imply deceleration.  We do not find significant asymmetry with respect to the outflow axis, which means that rotation around the outflow axis is insignificant.  If the outflow rotates fast, we will find a velocity gradient perpendicular to the outflow axis \citep[see, e.g.,][for an example of rotating outflows]{matsushita23}. \twelveco\ shows a larger and smoother deviation from the systemic velocity (note that the color scale of \twelveco\ is different from those of \hhco\ and \cs).
Still we find a similar decrease in the weighted velocity between shells 2 and 3 in \twelveco.  All the lines show the northwestern hole of a slow (or reverse) flow at a distance of $\sim 7\arcsec$  ($\sim 1000$ au) from the protostar.

Figure~\ref{fig:pv} shows the position velocity diagrams for \hhco\ and \cs\ along the outflow axis shown in the left panel of Figure~\ref{fig:moment1}a. The vertical dotted lines denote the locations of shells 1, 2, and 3.  They show a red component in addition to a blue component in the interval between the protostar and shell 3. The blue component shows a steep velocity gradient suggesting acceleration and terminates at shell 1. The red component appears around 1\arcsec\ ($\sim$150 au) from the protostar. The red component of \cs\ looks to continue beyond shell 2 while changing the velocity and intensity. We find another steep velocity gradient starting from shell 1.  Existence of the two components imply either diverging or converging flow along the line of sight, i.e., in the direction perpendicular to the outflow axis.
The velocity gradient is more likely to be due to acceleration/deceleration in the outflow rather than the initial velocity at the time of ejection. 

\cite{okoda25} derived the temperature and column density of H$_2$ molecules from the JWST MIRI MRS observations, which covers a $\sim 3\arcsec - 5 \arcsec $ region around the protostar (see their Figures 4a and 4b for the temperature and column density, respectively). The temperature is as high as $\sim 1300$~K near the protostar and decreases with the distance below 1000~K at shell 1.  The column density is as high as $ N \simeq 10 ^{20}~{\rm cm}^{-2} $ around the protostar and decreases to $ N \simeq 10 ^{19.5}~{\rm cm}^{-2} $ near shell 1. 

\begin{figure*}
\includegraphics[width=0.98\textwidth]{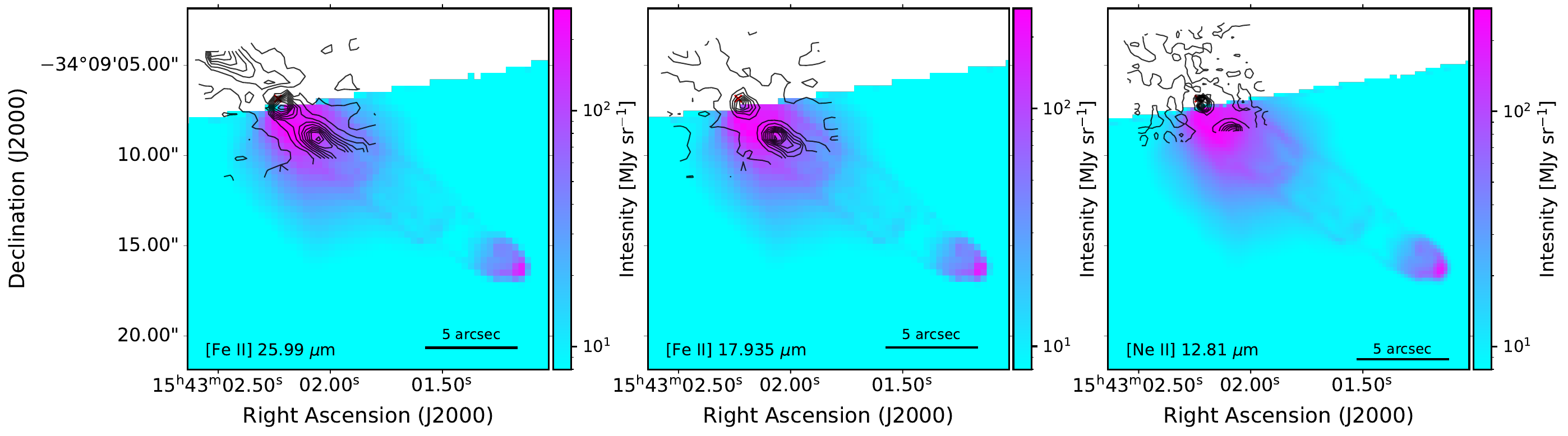}
\caption{Contours show intensities of (left) [\ion{Fe}{2}] 25.99 $\mu$m, (center) 17.935 $\mu$m, and (right) [\ion{Ne}{2}] 12.81 $\mu$m taken with JWST \citep{yang22}. The color denotes MIRI image taken with F560W filter on JWST. \label{fig:yang+22_fig11}}
\end{figure*}

Infrared emission from ionized elements supports the idea that the outflow gas is hot.  Figure \ref{fig:yang+22_fig11} show
continuum-subtracted intensity of [\ion{Fe}{2}] 25.99 $\mu$m, 17.935 $\mu$m, and [\ion{Ne}{2}] 12.81 $\mu$m.
 The blue background denotes MIRI F560W image taken with JWST \citep{yang22}. 
[\ion{Fe}{2}], [\ion{Ne}{2}], and [\ion{Ne}{2}] emissions are strong around the protostar and between the first and second shells.  In other words, they have a gap near the first shell, which is consistent with the temperature derived from H$_2$ line emissions. These ionized elements also show the substructure correlated with the shell structure. 
Differences among the line emissions maps are mostly due to the wavelength dependent angular resolutions. 

The temperature decrease with the distance from the protostar is not likely to be radiative cooling.  If the thermal energy is lost by radiative cooling, we need another heating source for the hot region between shells 1 and 2.  It is likely to be due to adiabatic expansion, which is consistent with the decrease in the column density and the increase in the blue-ward shift in the emission line. The adiabatic expansion converts thermal energy into a kinetic one, which can be used to heat up again the gas through a shock wave. 

\section{Underexpanded Jet}

A high pressure gas emerging from a nozzle evolves into an underexpanded jet. 
The study of such a jet has a long history. It can be traced back to the 19th century.  A review by \cite{franquet15} summarizes the history  and current knowledge. Though many numerical simulations have been done for astronomical jets and outflows, most of them are focused on the launch and collimation.  Some simulations are focused on the internal structure, i.e., the formation of the Mach disk \citep[see, e.g.,][]{norman82,komissarov98,smith22,smith24} and related structure.  However, industrial jets have been studied much more intensively and the theory is verified by experiments and well established.  Thus, we refer to \cite{franquet15} in the following since other reviews show essentially the same results.

In aeronautical science, the jet is divided into three zones: (1) the nearfield zone, (2) the transition zone, and (3) the farfield zone \citep[see, e.g.,][]{franquet15}. The nearfield zone is further divided into two components: the core and the mixing layer. The gas expands adiabatically in the core, where the pressure, density, and temperature decreases with distance from the end of nozzle. The pressure gradient accelerates the ejected gas to be supersonic.  The supersonic jet undergoes a shock and gets compressed to be hot and dense. The structure made by the shock is called the Mach disk. The jet between the nozzle end to the Mach disk is named ``barrel'' from its appearance (see Figure~\ref{fig:barrel-fig}). The jet width is controlled by the pressure balance between the jet and the ambient gas.
The hot dense gas ahead of the Mach disk is accelerated again to form subsequent substructures along the jet axis.

The morphology and structure of the jet depends mainly on the pressure ratio, $ P _0/ P _\infty $, where $ P _0 $ and $ P _\infty $ denote the initial pressure of the injected gas and the ambient gas pressure, respectively. The distance to the Mach disk ($ L _{\rm m} $) is proportional to the diameter of the nozzle ($ D _j$). \cite{crist66} derived an approximate formula,
\begin{eqnarray}
\frac{L _{\rm m}}{D _{\rm j}} & = & \sqrt{\displaystyle \frac{P _0}{2.4 \, P _\infty}} , \label{DmDj} 
\end{eqnarray}
analytically. \cite{young75} discussed that the coefficient in Equation (\ref{DmDj}) should be a function of the specific heat ratio, though the dependence is weak. Equation (\ref{DmDj}) is valid for a wide range of $ P _0/ P_\infty $. \cite{hatanaka12} examined the effects of the nozzle shape on the aspect ratio. Equation (\ref{DmDj}) gives a fairly good fit in the range $ 14 \la P _0/P _\infty \la 80 $ when the nozzle is either straight or divergent. Equation (\ref{DmDj}) overestimate $ D _m/D _j $ a little when the nozzle is straight. We can expect Equation (\ref{DmDj}) gives a good estimate for an astronomical jet though it does not emerge from a nozzle.  

The pressure ratio controls the ratio between the diameters of the Mach disk and jet nozzle. \cite{addy81} derived an empirical formula
\begin{eqnarray}
\frac{D _{\rm m}}{D _{\rm j}} & =& 0.36 \sqrt{\displaystyle \frac{P _0}{P _\infty} - 3.9 }  , \label{DmDj2}
\end{eqnarray}
for smooth convergent nozzle, where $ D _{\rm m} $ denotes the diameter of the Mach disk.  The ratio depends only slightly on the nozzle shape in particular when the pressure ratio is high.  

When the pressure ratio is high, the jet is variable. Though the flow in the core is laminar, the flow beyond the Mach disk is turbulent. See Figure 3 of \cite{yu13} shows the instantaneous and time-averaged images of the jet for $ P _0/ P _\infty = 10$.

We can evaluate the acceleration of the jet gas applying the Bernoulli's theorem. When the flow is stationary adiabatic in the barrel, the sum of the kinetic energy and specific enthalpy 
\begin{eqnarray}
H & = & \frac{1}{2} v ^2 + \frac{\gamma P}{(\gamma -1) \rho } = \frac{1}{2} v _0 ^2 + \frac{\gamma P _0}{(\gamma - 1) \rho _0},  \label{enthalpy} 
\end{eqnarray}
is constant along the stream line,
where $ v $, $ \rho $, and $ \gamma $ denote the velocity, density, and specific heat ratio, respectively. 
Given the initial velocity is much lower than the sound speed ($ | v _0 | \ll \sqrt{\gamma P _0}/{\rho _0} $, the jet becomes transonic when the temperature reaches $ T = 2 T _0/(\gamma + 1) $.  
The specific enthalpy is evaluated to be
\begin{eqnarray}
\frac{\gamma P}{(\gamma - 1) \rho} & = &
\left( 5.0~{\rm km~s}^{-1} \right) ^2 
\left( \frac{T}{2000~{\rm K}} \right) , \label{velocity}
\end{eqnarray}
for a molecular gas having the mean molecular weight, $ \mu = 2.3~m _{\rm H} $.  Equation (\ref{velocity}) provides an estimate for the gas velocity accelerated by the pressure.   

\section{Hydrodynamic Model}

\subsection{Model Setting}

We make a hydrodynamic model to reproduce the shell structure seen in \iras. For simplicity we assume that the outflow source maintain high temperature and high density . We ignore gravity, magnetic field, radiative cooling/heating processes for simplicity.  

 We use the cylindrical coordinates, $(r, \varphi, z )$ and assume the symmetry in the $ \varphi $-direction ($\partial/\partial \varphi = 0 $). Thus, the density ($ \rho $), pressure ($P$), and velocity ($ \mathbf{v} = (v _r, v _z)$) are functions of $ (r, z, t) $, where $ t $ denotes the time, which is set to be $ t = 0 $ at the initial stage, i.e., the onset of the outflow ejection.  
 The gas is assumed to be an ideal gas having the mean molecular weight $ \mu = 2.3  $ and the specific heat ratio, $ \gamma = 7/5 $. Then the pressure and density are expressed as
 \begin{eqnarray}
 P & = & n k_{\rm B} T , \\
 \rho & = & \mu m _{\rm H} n ,
 \end{eqnarray}
 where $ k _{\rm B} $ and $ m _{\rm H} $ denote the Boltzmann constant and the mass of hydrogen atom, respectively. We use the number density ($ n $) and temperature ($ T $) instead of $ P $ and $ \rho $ when we show our numerical results. 
  
 We assume a geometrically thin gas disk placed at $ z = 0 $ and consider the half space above the disk plane ($z >0$).  The half space is occupied initially by an ambient gas of which density and pressure are $ \rho _\infty $ and $ P _\infty $, respectively. The ambient gas is at rest at the initial stage ($ t = 0 $). 

We use the finite volume method in our numerical simulations.  The spatial resolution, i.e., the numerical cell width is uniform at $ \Delta r $ in the main part of the computation domain,
$ 0 \le r \le N _r \Delta r $ and $ 0 \le z \le N _z \Delta z $, where $ N _r $ and $ N _z $ are 780 and 7560, respectively. The cell width increases exponentially beyond $ r > N _r \Delta r $ and $ z > N _z \Delta r $. The cell width is larger by a factor of 1.25 compared to that of the inner adjacent cell. We placed 40 cells in the $ r $- and $ z $-directions outside the main part. Thus the outer boundary is located very far from the main part. We placed a fixed outer boundary at $ r = 2.47 \times 10 ^4~{\rm au} $ and $ z = 2.47 \times 19 ^4~{\rm au}$ but no disturbance reaches the outer boundary. 

We set an outflow engine as a boundary condition on $ z = 0 $ plane. The nozzle of the jet is assumed to be a circle having the radius, $ R _0$.
The pressure is fixed to be
\begin{eqnarray}
P (r, z = 0) & = & 
\begin{cases}
P _0 & (r \le R _0 - 3 \Delta r ) \\
P _\infty & (z \ge 0) \\
\end{cases} .
\end{eqnarray}
The pressure decreases from $ P _0 $ to $ P _\infty $ in the $ R _0 - 3 \Delta r < r < R _0 $ at a constant rate.
The density and velocity are fixed to be
\begin{eqnarray}
\rho (r, z < 0 ) & = &
\begin{cases}
\rho _0 & (r \le R) \\
\rho _\infty & (r > R) \\
\end{cases} , \\
v _r (r, z < 0 ) & = & 0 \\
v _z (r, z < 0 ) & = &
\begin{cases}
v _{\rm jet} & (r \le R) \\
0 & (r > R) \\
\end{cases} ,
\end{eqnarray}
As summarized in Table \ref{tab:modelP} we set the initial flow to be 
subsonic  ($ v _{\rm jet}/\sqrt{\gamma P _0/\rho _0} = 0.27 $).
Remember that the supersonic collimated
outflow was given as the initial condition in the simulations of \cite{norman82}, \cite{smith22}, and others.
Our simulations include initial acceleration and collimation though we assume hot dense gas and disk
as a source of acceleration and collimation, respectively.

We use the number density, $ n _0 = \rho _0 /(\mu m _{\rm H}) $ and $ n _\infty = \rho _\infty /(\mu m _{\rm H}) $, instead of $ \rho _0 $ and $ \rho _\infty $ to specify the initial density distribution. We also use $ T _0 = P _0 /(k _{\rm B} n _0) $ and $ T _\infty = P _\infty / (k _{\rm B} n _\infty) $ to define the pressure distribution. We solve the hydrodynamical equations numerically with the second order accurate scheme which can capture a shock wave without spurious oscillations. We employ a flux split method based on approximate Riemann solution to avoid negative density and pressure.  Our numerical code has been proved to reproduce a strong shock wave by solving the shock tube problem. It has been applied to gas accretion to a gas disk rotating around a protostar \citep{hanawa21,hanawa22,hanawa24}.

\subsection{Canonical Model}

First we show model A in which the model parameters are taken to be $ n _0 = 10 ^7~{\rm cm}^{-3} $, $ n _\infty = 5 \times 10 ^6~{\rm cm}^{-3}$, $ T _0 = 2000~{\rm K} $, and $ T _\infty = 100~{\rm K}$. Then the pressure ratio is set to be $ P _0/P _\infty = n _0 T _0 / (n _\infty T _\infty) = 40 $.  The jet is assumed to emerge from the circle of $ R = 50~{\rm au} $ with the initial velocity $ v _{\rm jet} = 0.845~{\rm km~s}^{-1}$, which is much smaller than the sound speed at 2000~K, 3.18~km~s$^{-1}$.  The spatial resolution is $ \Delta r = 0.714~{\rm au} $ in the main part of the computation box of $ 0 \le r \le 557~\rm{au}$ and $ 0 \le z \le 5400~{\rm au} $. The jet emerging circle is resolved with $ R/\Delta r = 70 $ numerical cells. 

Figure~\ref{2D_09w-75} shows model A by a snap shot at $ t = 7.47 \times 10 ^3 $~yr. The top and second panels show the density ($ n $) and the temperature ($T$) in the logarithmic scale by color. The third and fourth panels show the velocity $(v _r, v _z )$. Each panel shows also the area of $ r < 0 $, which is flipped over from that of $ r > 0 $, for comparison with \iras. All the panels show the Mach disk at $ z  \simeq 300~{\rm au}$.  Though the flow is laminar in the barrel ($ z \la 300~{\rm au} $), it is turbulent in the region beyond the Mach disk. The flow is variable as shown in the animation associated with Figure~\ref{2D_09w-75}.  Still we find a semi-periodic change in the velocity, $ (v _r, v _z) $. This change is due to the pressure balance between the outflow and ambient gases. The longitudinal velocity is negative ($ v _z < 0 $) near $ z = 1000~{\rm au} $. This substructure evolves into the second shell. The outflow compresses the ambient gas ahead of it through a shock.  The shock compressed gas has the highest density and a moderate temperature of several hundreds K.   

\begin{figure}
\plotone{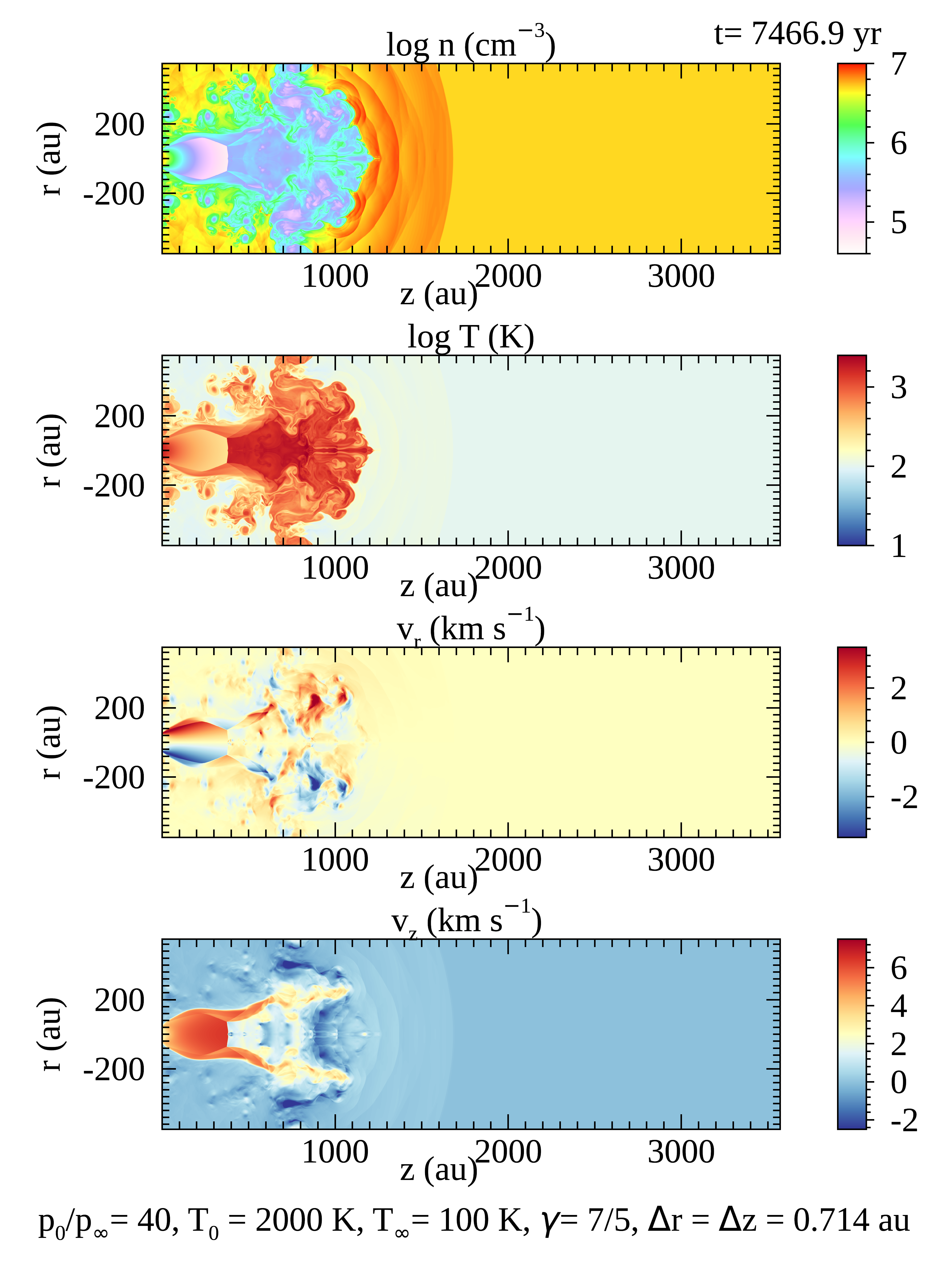}
\caption{Each panel shows the number density ($n$), temperature ($T$), and velocity ($ v _r $, $ v _z $) at $ t = 7.47 \times 10 ^3 {\rm yr} $ in model 40. The time evolution of model A is shown in the associated movie. The animation shows that the gas emerges from the left and the head proceeds rightward. The outflow is turbulent and associated with shock waves. The animation shows the sequence from zero to 443.98 years. The duration of the animation is 18 seconds.\label{2D_09w-75}}
\end{figure}

\begin{table}[h]
    \begin{center}
    \begin{tabular}{cccccc}
    \hline
      model   & $ P _0/P _\infty $ & $ n _0 / n _\infty $ &$ T _0 $ & $T _\infty $ & $v _{\rm jet}$\\
      \hline
     A    &  40 & 2.0 & 2000~K & 100~K & 0.85~km~s$^{-1}$\\  
     B   &  62.5 & 3.125 &  2000~K & 100~K & 0.85~km~s$^{-1}$\\ 
     C    &  40 & 0.4  &2000~K & 20~K & 0.85~km~s$^{-1}$\\ 
     \hline
    \end{tabular}
    \end{center}
    \caption{Model Parameters.}
    \label{tab:modelP}
\end{table}

Figure \ref{2D_09w-152} is the same as Figure~\ref{2D_09w-75} but shows a later stage at $ t = 1.498 \times 10 ^4 {\rm yr} $.
The outflow head reaches $ z \sim 2200~{\rm au} $ and the density is highest in the shock compressed gas ahead of it. 
The semi-periodic structure is prominent in the velocity. 
\begin{figure}
\plotone{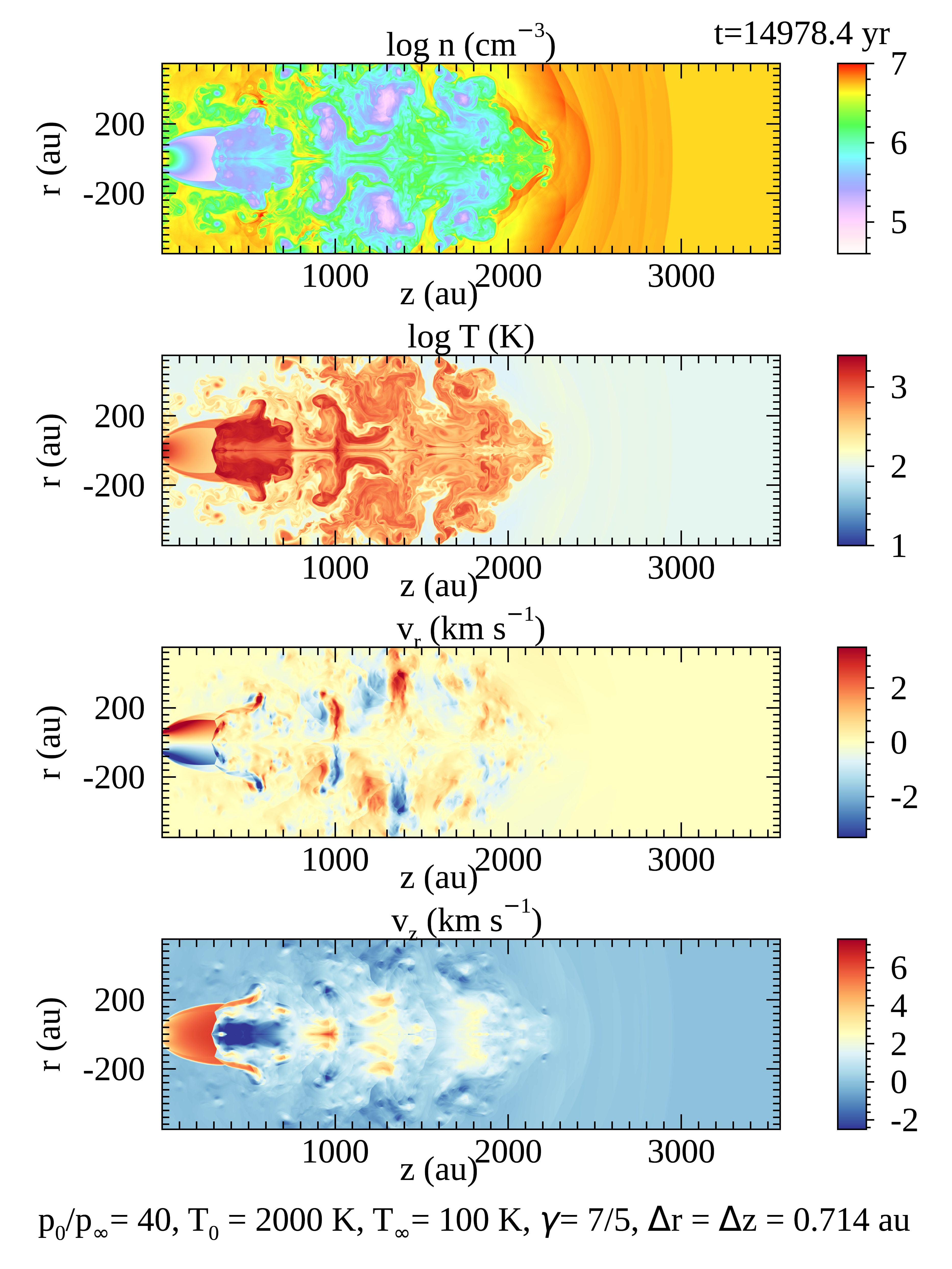}
\caption{The same as Figure~\ref{2D_09w-75} but for the stage at $ t= 1.498 \times 10^4~{\rm yr} $. \label{2D_09w-152}}
\end{figure}

Figure \ref{2D_09w-152Mach} shows the barrel structure at $ t = 1.498 \times 10 ^{4}~{\rm yr} $. The upper and lower panels
denote the density and temperature distributions by color, respectively. The arrows overlaid on the upper panel denote the velocity.
As expected, the structure is essentially the same as that illustrated in Figure~\ref{fig:barrel-fig}. A hot dense gas flows between the jet boundary and the oblique shock wave. The Mach disk is bent at this stage.  See the animation associated with Figure~\ref{2D_09w-152Mach} for the formation and evolution of the Mach disk.

\begin{figure}
\plotone{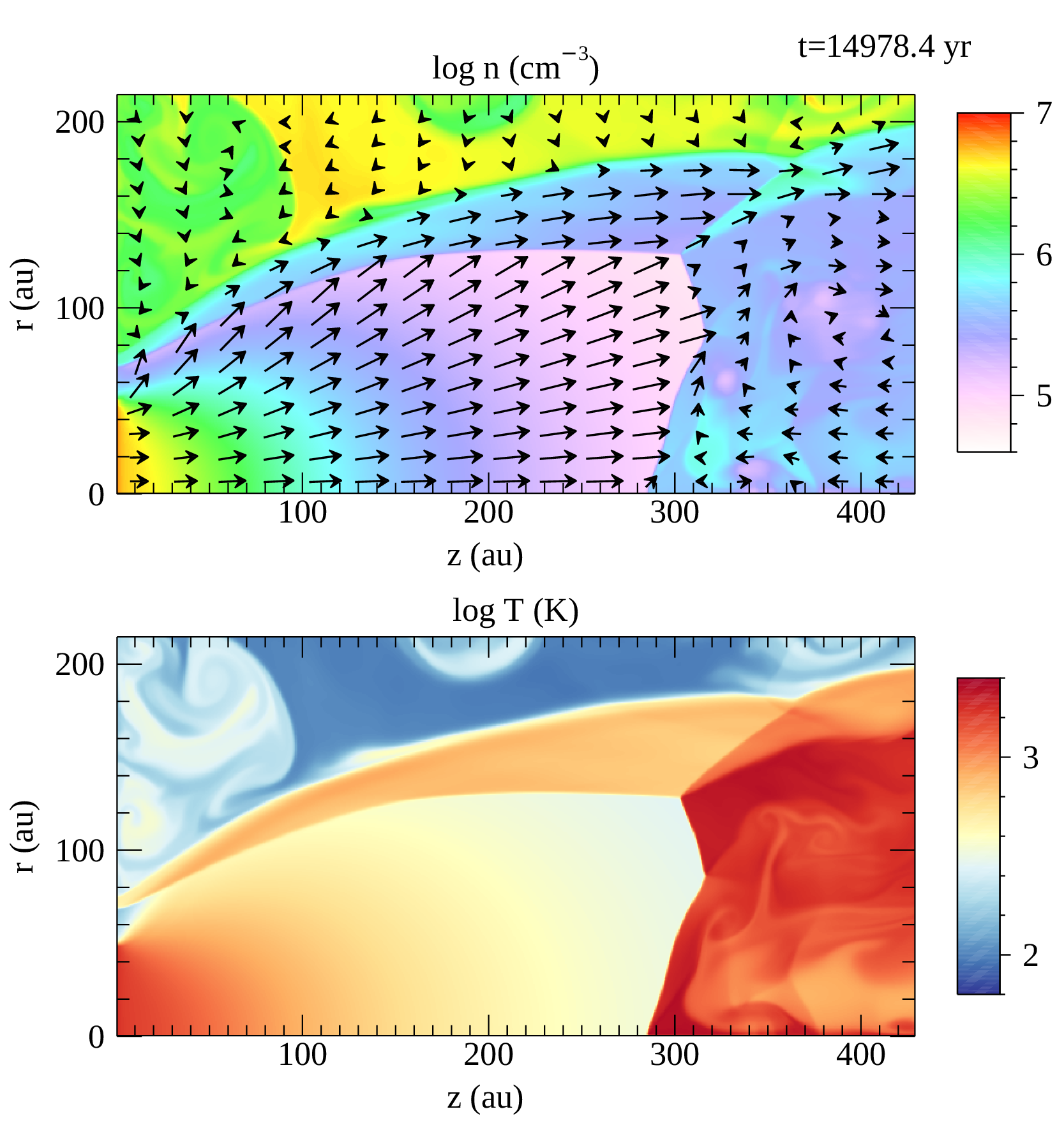}
\caption{The flow structure in the barrel in model A at $ t= 1.498 \times 10^4~{\rm yr} $. The associated movie shows the fluctuation of the Mach disk. The first barrel is persistent but fluctuates. The animation shows the sequence from zero to 29455.5 years. The duration of the animation is 12 seconds. \label{2D_09w-152Mach}}
\end{figure}

Figure \ref{2D_09w-vz} shows the longitudinal velocity ($ v _z $) at the stage shown in Figure~\ref{2D_09w-152} as a function of $ z $, i.e., the distance from the nozzle.  The dark blue curve denotes $ v _z $ on the $ z $-axis while the light blue curve shows the maximum of $ v _z $ at a given $ z $.  The jet flow is not straight as shown in the bottom panel of Figure \ref{2D_09w-152}.   The longitudinal velocity has a large jump at several places, i.e., at the shock fronts. The Mach number, the ratio of the flow velocity and sound speed, reaches 4.75 and drops suddenly at $z = 293~{\rm au}$, i.e., just in front of the Mach disk.  The Mach number exceeds unity at $ z = 7.5~{\rm au} $ and the flow is supersonic from there to the Mach disk. The jet flow is off-axis in the range of $ 300~{\rm au} \la z \la 800~{\rm au} $.  The flow is reverse ($ v _z < 0 $) around the axis in this range.  Remember that Figure~\ref{fig:pv} shows two velocity components along the outflow axis. The blue-shifted may be the off-axis flow along our line of sight while the red-shifted one may be the flow near the outflow axis.

\begin{figure}
\plotone{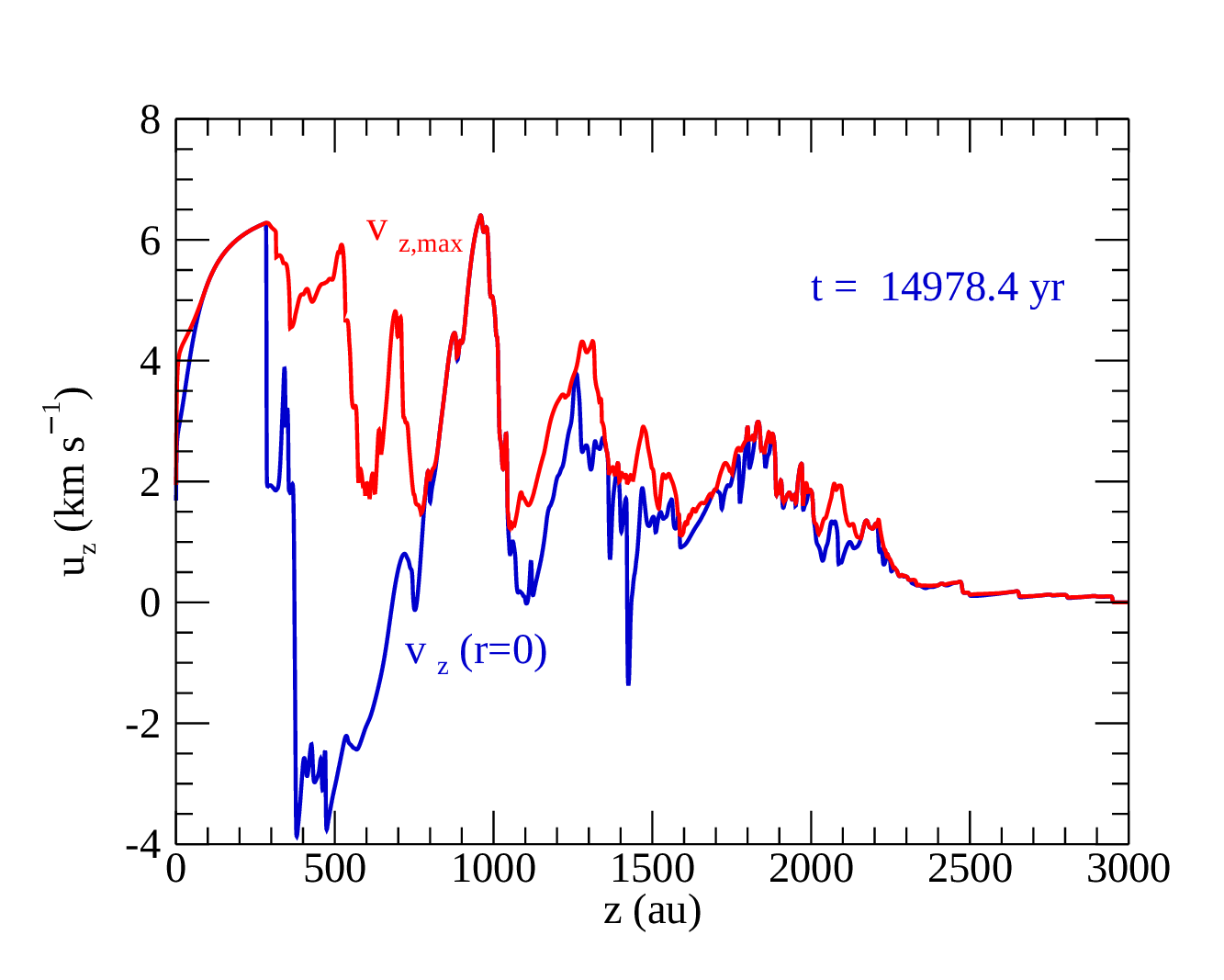}
\caption{The longitudinal velocity $(u _z)$ at $ t = 1.498 \times 10 ^4~{\rm yr} $ in model A. \label{2D_09w-vz}}
\end{figure}

Figures \ref{2D_09w-rho} and \ref{2D_09w-temp} show the density and temperature distributions on the $z$-axis at several epochs, respectively. The density and temperature decrease steeply as soon as the gas emerges from the nozzle.  Remember that the jet gas has $ n _0 = 10 ^7~{\rm cm} ^{-3} $ and $ T _0 = 2000~{\rm K} $ at $ z = 0 $. Also the pressure decreases steeply near the nozzle. Since the pressure ratio is very high $ P _0/P _\infty = 40 $, the jet gas expands rapidly and adiabatically. The flow is supersonic in the barrel. The density and temperature decrease below $ 10 ^5~{\rm cm}^{-3} $ and $ 350~{\rm K} $, respectively, on the upstream side of the Mach disk. They rise again to $ 5 \times 10 ^5~{\rm cm} ^{-1} $ and $\sim 2000~{\rm K} $ on the downstream side of the Mach disk by the shock compression. This dense hot gas expands again to accelerate the gas ahead. Figures \ref{2D_09w-rho} and \ref{2D_09w-temp} show the laminar flow in the barrel $(z < 300~{\rm au}) $ is stationary while the flow beyond the Mach disk is variable.  The head of the jet proceeds at almost constant rate.

\begin{figure}
\plotone{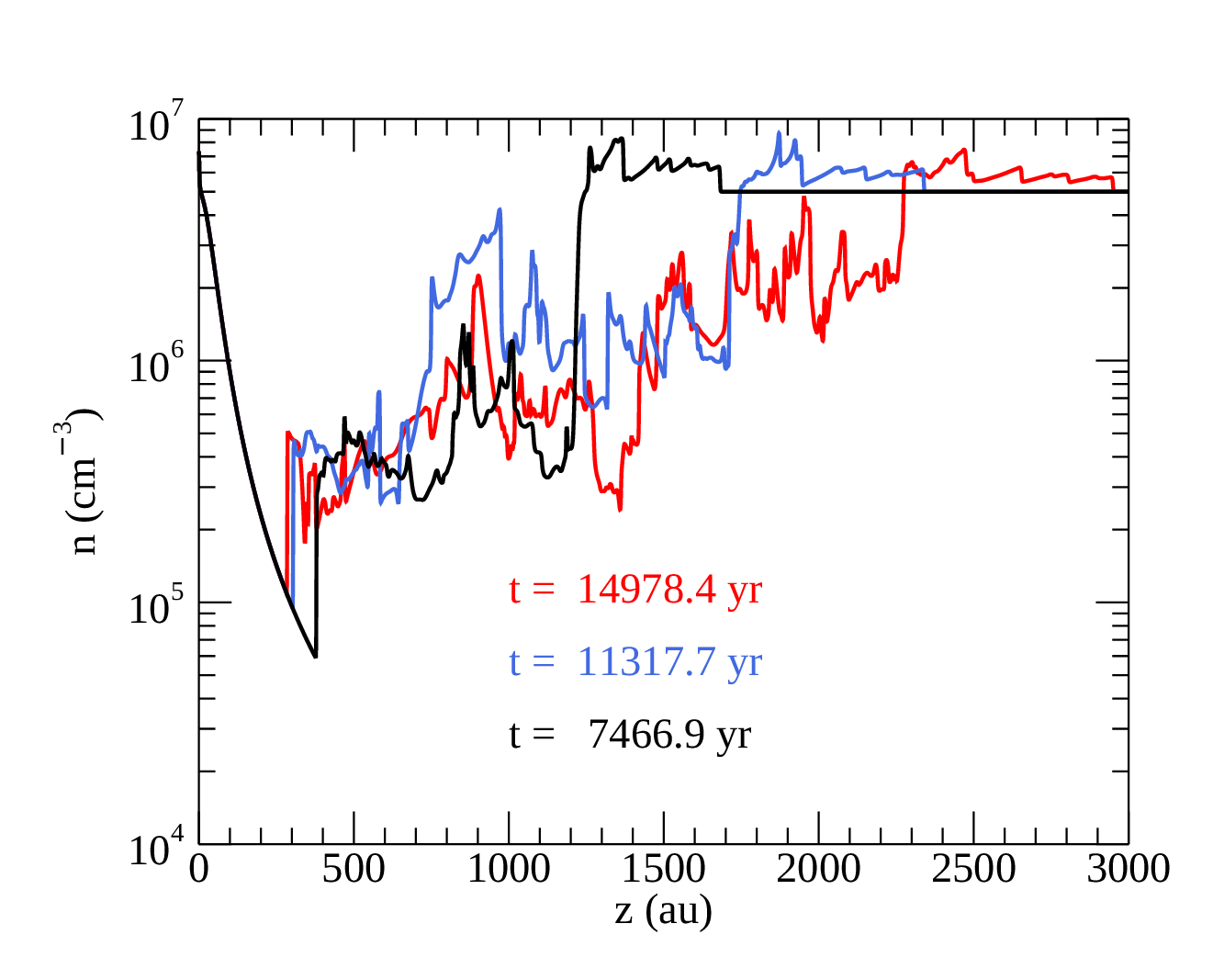}
\caption{The density distribution on the $z$-axis in model A at several epochs. \label{2D_09w-rho}}
\end{figure}

\begin{figure}
\plotone{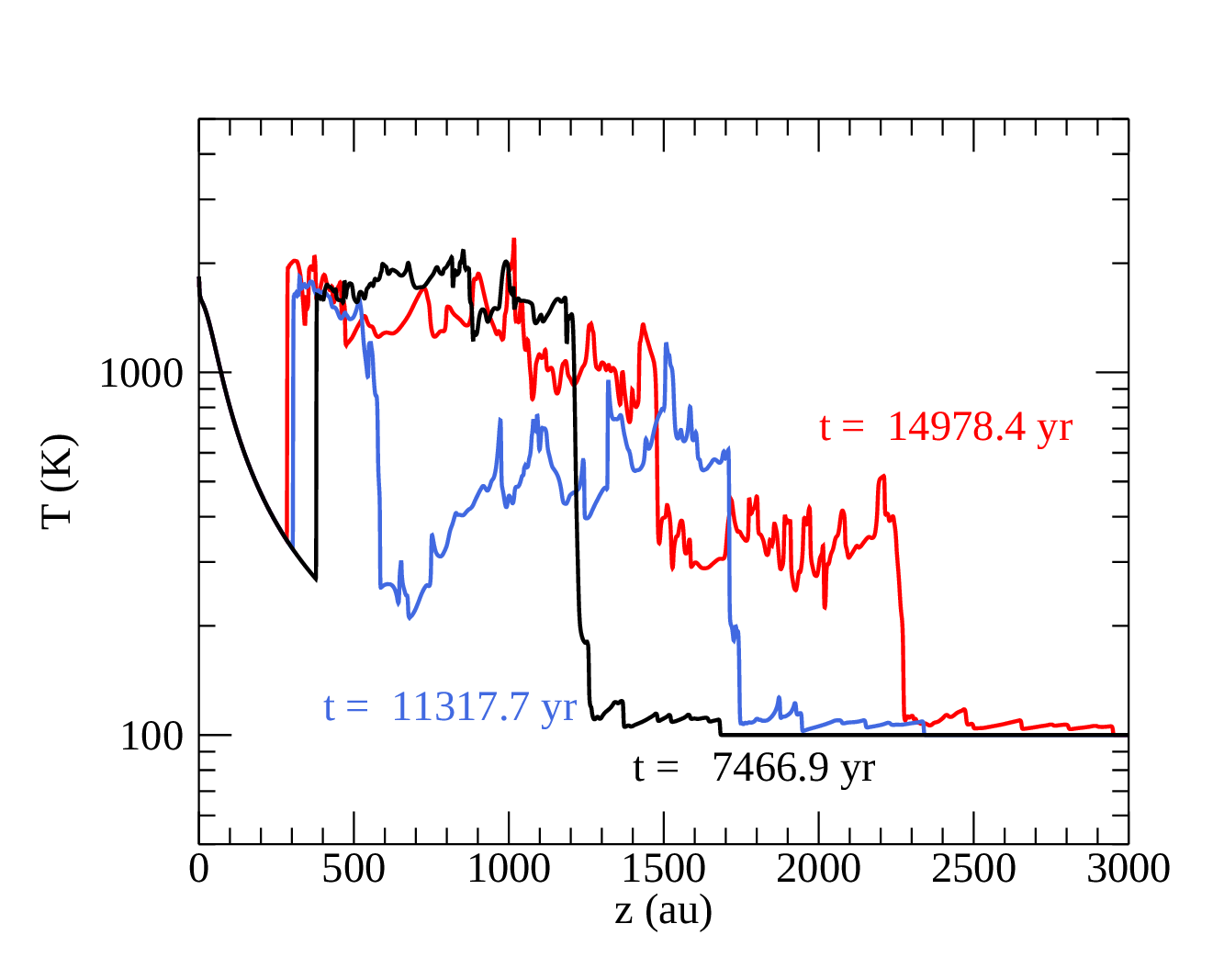}
\caption{The temperature distribution on the $z$-axis in model A at several epochs. \label{2D_09w-temp}}
\end{figure}

We measure the width and height of the barrel from the velocity profile. The tangential and longitudinal components of the velocity ($ v _r $ and $ v _z $) are discontinuous on the side of the barrel and at the Mach disk, respectively.  We measure the radius where $ v _r $ turns to decrease at a given height, $ z $ and define it as the local barrel radius. The barrel diameter ($D _b$) is defined as twice the maximum barrel radius.  We define the barrel height as the place where $ v _z $ turns to decrease on the axis ($r = 0$). Figure \ref{2D_09w_barrel} shows the width and height of the barrel in model A. The height is slightly smaller than expected from Equation (\ref{DmDj}), 408~au. The height is $\sim 20 \%$ larger than the width, though both of them are oscillating substantially. The diameter of the Mach disk is 210 au at the stage shown in Figure~\ref{2D_09w-75}.

\begin{figure}
\plotone{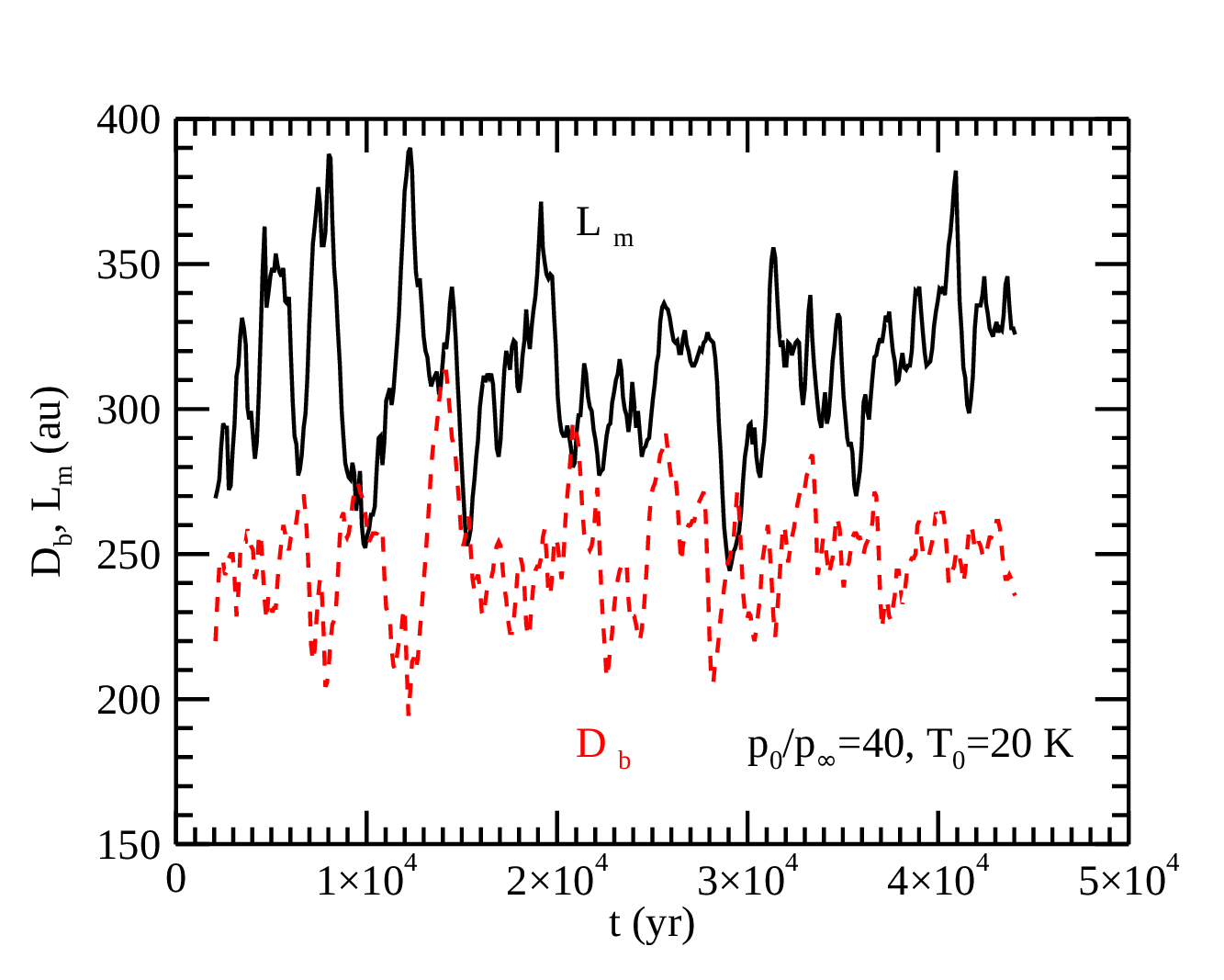}
\caption{The time evolution of the barrel width (red) and height (black) in model A. \label{2D_09w_barrel}}
\end{figure}

\subsection{Higher Pressure Ratio Model}

We show model B to examine the effects of a higher pressure ratio.  The initial state of model B is the
same as that of model A except for the pressure ratio (see Table~\ref{tab:modelP}), i.e., the density ratio.  The initial density of the injected gas is set to be $ n _0 = 10 ^7~{\rm cm} ^{-3}$ for easier comparison with model A. Accordingly, the ambient density is lower in model B.
The temperature ratio remains at $(T _0/T _\infty = 20 $).  
Figure~\ref{2D_11j_154} is the same as Figure~\ref{2D_09w-152} but for the stage 
at $ t = 1.494 \times 10 ^4~{\rm yr} $ in model B. As shown in the animation associated with Figure~\ref{2D_11j_154},
models A and B are qualitatively similar.  

\begin{figure}
\plotone{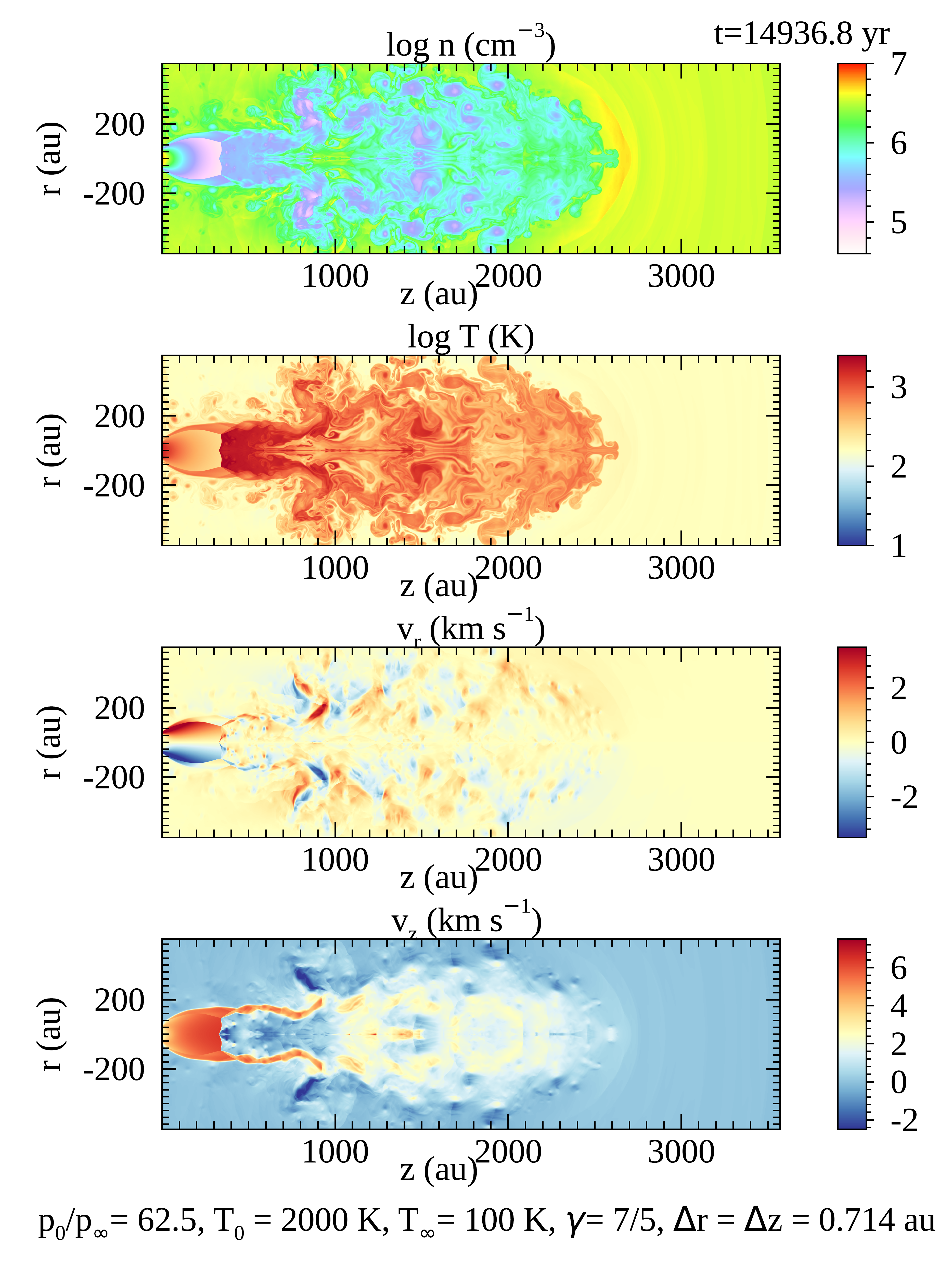}
\caption{The same as Figure~\ref{2D_09w-152} but for model B at $ t =  1.494 \times 10 ^4~{\rm yr} $. The associated animation shows the development of the outflow in model B.  The outflow is turbulent and associated with shock waves. The first barrel is persistent. The animation shows the sequence from zero to 20874.4 years.   The duration of the animation is 9 seconds.\label{2D_11j_154}}
\end{figure}

Figure~\ref{2D_11j_barrel} is the same as Figure~\ref{2D_09w_barrel} but for model B. 
The first barrel of model B has a larger height and width than that of model A.
The aspect ratio, the height over the width is larger in model B, i.e., 
the first barrel is more slender in model B.  Both the height and width of the outflow are larger for a higher pressure ratio.  But, the dependence of the width on the pressure ratio is weaker than that of the height.
Another difference is the density contrast between the outflow and ambient gas. The ambient gas is less dense in model B.

\begin{figure}
\plotone{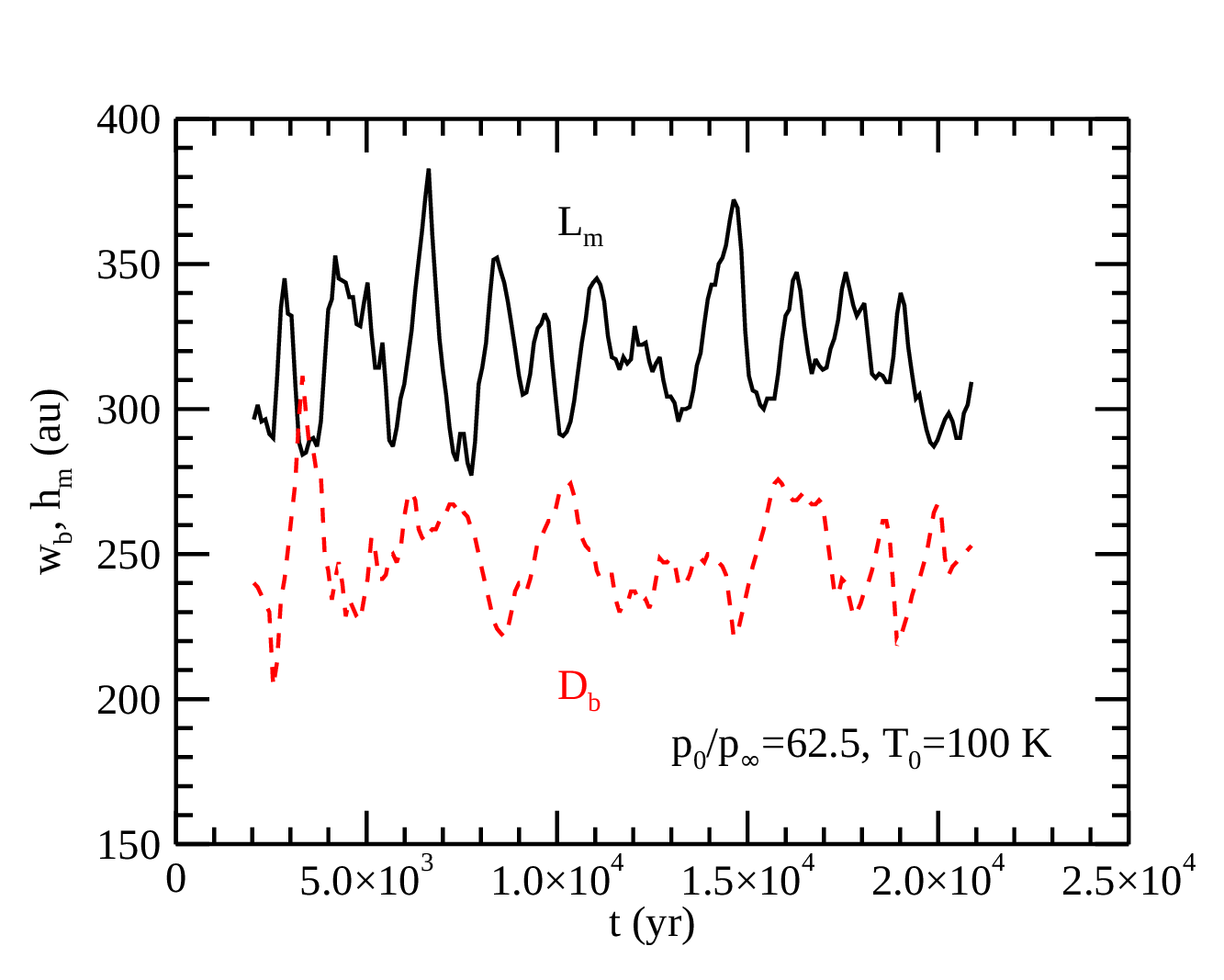}
\caption{The same as Figure \ref{2D_09w_barrel} but for model B. \label{2D_11j_barrel}}
\end{figure}

\subsection{Cold Ambient Gas Model}

We construct model C in which the ambient gas is set to be cold ($ T _\infty = 20~{\rm K} $).  The pressure ratio 
is set to be $ P _0/ P _\infty = 40 $ for comparison with model A.  Since the temperature contract is 
$ T _0 / T _\infty = 100 $, the ambient gas is 2.5 times denser than that of the injected gas.  

Figure \ref{2D_11d_360} shows the stage at $ t = 3.545 \times 10 ^4~{\rm yr} $ in model C.
Model C is also qualitatively similar to models A and B.  The outflow shows the barrel 
and subsequent semi-periodic structure.  However, the outflow is less prominent since the ambient gas is
much denser.  Also the outflow propagates more slowly than in models A and B.  The slow propagation
is also due to the dense ambient gas.  The larger inertia of the ambient gas decelerates the outflow propagation 
substantially. This density dependence is similar to that found by \cite{smith22} though their initial
set-up is different.

\begin{figure}
\plotone{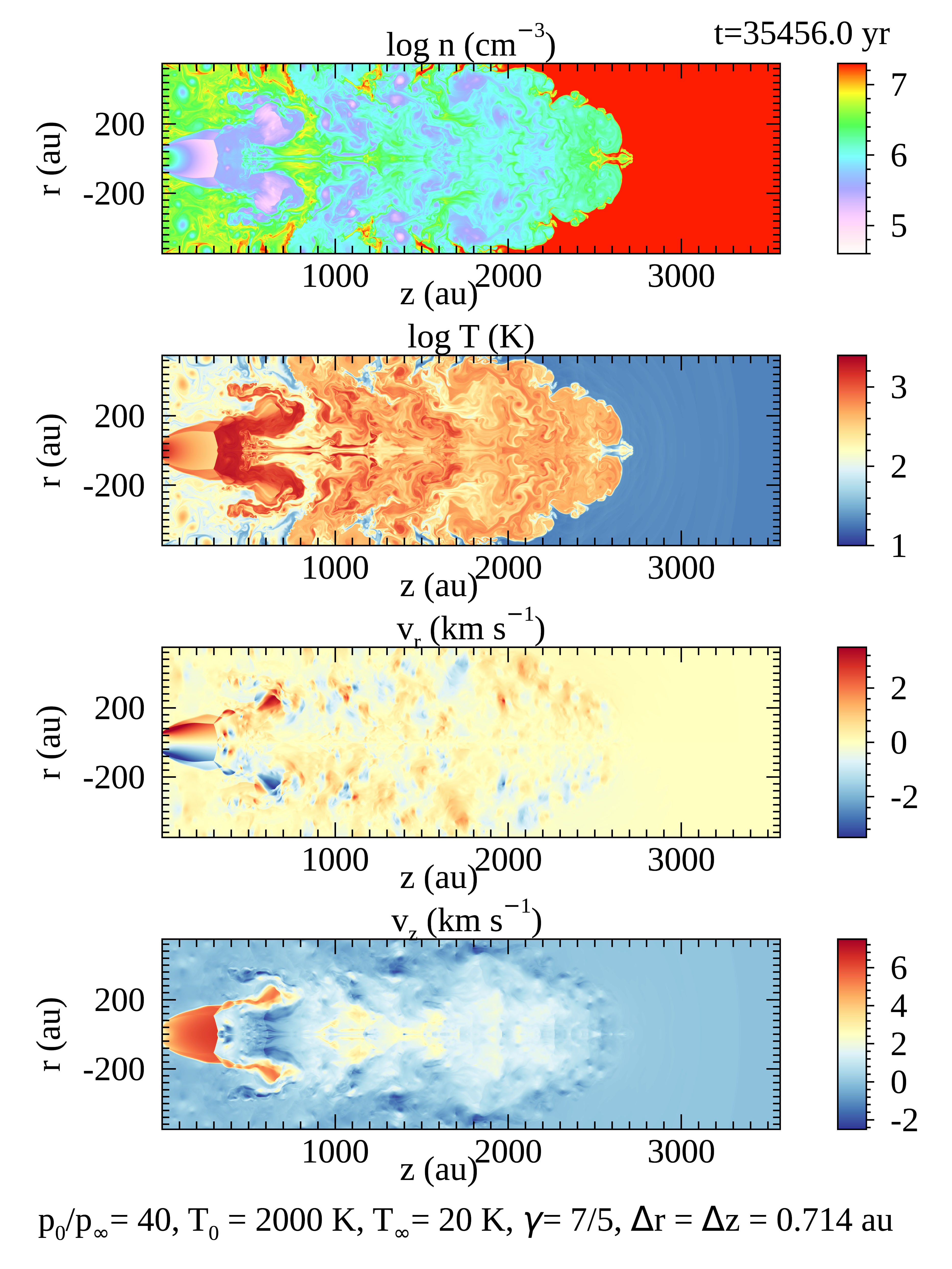}
\caption{The same as Figure~\ref{2D_09w-152} but for model C at $ t =  3.545 \times 10 ^4~{\rm yr} $. The associated animation shows the development of outflow in model C.  The propagation is substantially slower in model C. The animation shows the sequence from zero to 31289.8 years.  The duration of the animation is 18 seconds.\label{2D_11d_360}}
\end{figure}

\section{Discussions and Summary}

As shown in the previous sections, our simple hydrodynamic model explains several characters of the primary outflow in \iras.
First, it explains the shell structure in the intensity and that in the velocity.  Second, it explains the anti-correlation between the temperature and velocity. It also explains the magnitude of the outflow velocity. Equations (\ref{enthalpy}) and (\ref{velocity}) suggest that the outflow is accelerated to $ \sim 7~{\rm km~s}^{-1} $ if most of the initial thermal energy is converted into the kinetic energy at $ T _0 = 2000~{\rm K} $. The inferred value is consistent with the observations since the outflow is inclined from our line of sight. Fourth, our model shows a backward ($ v _z < 0 $) flow in the outflow, which may explain the red-shifted flow in the blue lobe.
Our model demonstrates that we need not invoke short time variability in the outflow activity to explain the shell structure.

Our model is an oversimplification in some regards. We assumed that the outflow source maintains constant pressure, density and temperature after a sudden onset.  However, the outflow activity is likely to decay after a short rise. The pressure ratio may be a little higher in the early phase of the primary outflow activity.  Uniform density in the source region is also an oversimplification.  Thus, we think that our model has limitations to explain the primary outflow of \iras.

Nevertheless, our model gives some constraints on a better model.  First, the ambient gas density should be low enough. The outflow gas is hot ($T > 1000~{\rm K}$) and should have a high pressure to accelerate the gas to be highly supersonic.  The density of the outflow gas decreases considerably during the acceleration.  If the initial density of the outflow is comparable or lower than that of the ambient gas, the outflow is decelerated by the inertia of the ambient gas (see model C). \iras\ experienced at least two earlier outflow activities \citep{2021ApJ...910...11O,sai24}.
The ambient gas around \iras\ can be blown off by the third or even earlier outflow.  \cite{redalelli19} estimated the gas density to be $ n _{{\rm H}_2} = 2.6 \times 10 ^4~{\rm cm}^{-3} $ around \iras\ from the radio and far infrared continuum emission observed with Herschel.  They derived the value by approximating the gas cloud to be a uniform sphere having the effective radius 0.1~pc.  Thus, this value should be taken as an average in the region of $ 2 \times 10 ^4 $~au around \iras.  If the gas density around the primary outflow is similarly low, the outflowing gas will have a higher density than the ambient gas if the outflowing gas has the initial density $ n _{{\rm H}_2} \approx 10^7~{\rm cm}^{-3} $. Second, the gas temperature should be as high as 2000~K around the protostar.  The outflow gas has a high temperature \citep[$ T > 1000~{\rm K} $,][]{okoda25} and substantial velocity.  This implies the outflow gas has a higher temperature before acceleration unless it is heated up during the acceleration.  Third, the source of the outflow should have a radial size of $ \sim 50~{\rm au} $ if the first shell located at 300~au from the protostar is due to the Mach disk.

Though JWST observed \iras, the measurement of the gas temperature is limited to a narrow region around the protostar \citep[see Figure 8 of][]{okoda25}, i.e., the field of view of the spectroscopic observation.  It is interesting to measure the temperature and density in the outflow beyond the second shell.  Similarly, ALMA observed only a limited portion of the primary outflow with the highest observation \citep{thieme23}.  It would be interesting to study the kinematics around shells 3 and 4 with ALMA.   

Also interesting is to study outflows showing similar substructures.  Herbig-Haro 211 (hereafter HH 211) is an example showing a similar substructure in the CO ($J=2-1$) \citep{gueth99}) and H$_2$ (2.1~$\mu$m) \citep{mccaughrean94} emission lines. Figure 1 of \cite{cabrit11} shows the correlation between the substructures.  
The infrared emission from HH 211 was further studied with Spitzer telescope \cite{tappe12} and JWST \citep{ray23}. \cite{ray23} identified that the 4.5 $\mu$m localized emissions are largely due to the CO fundamental rotational-vibrational transitions.  This means that the CO gas is hot ($\sim 1000~{\rm K} $) only in the narrow regions where the 4.5~$\mu$m emission is strong.  Other examples are CALMA 7 and DG Tau B.  \cite{plunkett15} discovered semi-regularly placed knots in the outflow from CARMA 7, a Class 0 source in the Serpens South star forming region.  They found 11 knots each in the red and blue lobes. 
\citep{devalon20} discovered shell structure in the outflow cavity of DG Tau B. The shells appear near the protostar in the widely open conical cavity.  We may find other examples from archival data of ALMA.

\begin{acknowledgements}
This work is supported by the MEXT/JSPS Grant-in-Aid from the Ministry of Education, Culture, Sports, Science, and Technology of Japan (20H00182, 20H05844, 20H05845, 20H05847, and 22K20389).  
Y.O., Y.-L.Y., and N.S. are grateful for support from the RIKEN pioneering project: Evolution of Matter in the Universe. 
Y. O. is supported by RIKEN Special Postdoctoral Researcher Program (Fellowships) and the Japan Society for the Promotion of Science (JSPS), Overseas Research Fellowship.
Japan Society for the Promotion of Science (JSPS), Overseas Research Fellowship.
This work is based in part on observations made with the NASA/ESA/CSA James Webb Space Telescope. The data were obtained from the Mikulski Archive for Space Telescopes at the Space Telescope Science Institute, which is operated by the Association of Universities for Research in Astronomy, Inc., under NASA contract NAS 5-03127 for JWST. These observations are associated with program \#2151.
The data were retrieved from the Mikulski Archive for Space Telescopes at the Space Telescope Science Institute, operated by the Association of Universities for Research in Astronomy (AURA), Inc., under NASA contract NAS 5-03127. 
The data presented in this paper were obtained from the Mikulski Archive for Space Telescopes (MAST) at the Space Telescope Science Institute. The specific observations analyzed can be accessed via DOI: \dataset[10.17909/wv1n-rf97]{https://doi.org/10.17909/wv1n-rf97} and \citet{2024ApJ...974...97S}.
This paper makes use of the following ALMA data: ADS/JAO.ALMA\#2013.1.00879.S and ADS/JAO.ALMA\#2018.1.01205.L. ALMA is a partnership of the ESO (representing its member states), the NSF (USA) and NINS (Japan), together with the NRC (Canada) and the NSC and ASIAA (Taiwan), in cooperation with the Republic of Chile. The Joint ALMA Observatory is operated by the ESO, the AUI/NRAO, and the NAOJ.
\end{acknowledgements}


\bibliography{Jet.bib}{}
\bibliographystyle{aasjournal}

\end{document}